 \documentclass[preprint]{aastex}

\newbox\grsign \setbox\grsign=\hbox{$>$} \newdimen\grdimen
\grdimen=\ht\grsign
\newbox\simlessbox \newbox\simgreatbox \newbox\simpropbox
\setbox\simgreatbox=\hbox{\raise.5ex\hbox{$>$}\llap
     {\lower.5ex\hbox{$\sim$}}}\ht1=\grdimen\dp1=0pt
\setbox\simlessbox=\hbox{\raise.5ex\hbox{$<$}\llap
     {\lower.5ex\hbox{$\sim$}}}\ht2=\grdimen\dp2=0pt
\setbox\simpropbox=\hbox{\raise.5ex\hbox{$\propto$}\llap
     {\lower.5ex\hbox{$\sim$}}}\ht2=\grdimen\dp2=0pt
\def\simgreat{\mathrel{\copy\simgreatbox}}
\def\simless{\mathrel{\copy\simlessbox}}

 \slugcomment{To appear in The Astrophysical Journal, v565 Jan 20, 2002 issue}

\shorttitle{A Chandra X-ray Study of Cygnus~A: III. The Cluster of Galaxies}
\shortauthors{Smith et al.}

\begin{document}

\title{A Chandra X-ray Study of Cygnus~A --- III. The Cluster of Galaxies}


\author{David A. Smith and Andrew S. Wilson\altaffilmark{1}}

\affil{Department of Astronomy, University of Maryland, College Park, MD
20742; dasmith@astro.umd.edu, wilson@astro.umd.edu}

\author{Keith A. Arnaud\altaffilmark{2}}
 
\affil{Laboratory for High Energy Astrophysics, NASA/GSFC, Code 662,
Greenbelt, MD 20771; kaa@genji.gsfc.nasa.gov}


\and

\author{Yuichi Terashima\altaffilmark{3} and Andrew J. Young}
 
\affil{Department of Astronomy, University of Maryland, College Park, MD
20742; terasima@astro.umd.edu, ayoung@astro.umd.edu}

\altaffiltext{1}{Adjunct Astronomer, Space Telescope Science
Institute, 3700 San Martin Drive, Baltimore, MD 21218;
awilson@stsci.edu}

\altaffiltext{2}{also Department of Astronomy, University of Maryland,
College Park, MD 20742}

\altaffiltext{3}{Institute of Space and Astronautical Science, 3-1-1
Yoshinodai, Sagamihara, Kanagawa 229-8510, Japan}

\begin{abstract}

We present an analysis of the \emph{Chandra} Advanced CCD Imaging
Spectrometer (ACIS) observation of the intracluster gas associated
with the cluster of galaxies surrounding Cygnus~A.  The dominant
gaseous structure is a roughly elliptical (presumably prolate
spheroidal in three dimensions) feature with semi-major axis $\simeq
1.\!^{\prime}1$ ($\simeq 100$~kpc). This structure apparently
represents intracluster gas which has been swept up and compressed by
a cavity inflated in this gas by relativistic material which has
passed through the ends of the radio jets.  The X-ray emitting gas
shows this prolate spheroidal morphology to $\simeq 1.\!^{\prime}2$
(110~kpc) from the radio galaxy, but is spherical on larger scales.
The X-ray emission from the intracluster gas extends to at least
$8^{\prime}$ ($\simeq 720$~kpc) from the radio galaxy, and a second,
extended source of X-ray emission (probably associated with a second
cluster of galaxies) is seen some $12^{\prime}$ ($\simeq 1$~Mpc) to
the NW of Cygnus~A.  The X-ray spectrum of the integrated intracluster
gas imaged on the S3 chip (dimensions $8^{\prime} \times 8^{\prime} =
720 \times 720$~kpc), excluding the contribution from the radio galaxy
and other compact sources of X-ray emission, has a gas temperature,
metallicity, and unabsorbed 2--10~keV rest-frame luminosity of
$7.7$~keV, $0.34$ times solar, and $3.5 \times 10^{44}$ erg s$^{-1}$,
respectively.

We have deprojected the X-ray spectra taken from 12 elliptical and
circular annuli in order to derive a run of temperature, metallicity,
density, and pressure as a function of radius.  The temperature of the
X-ray emitting gas drops from $\simeq 8$~keV more than $100$~kpc from
the center to $\simeq 5$~keV some $80$~kpc from the center, with the
coolest gas immediately adjacent to the radio galaxy.  ``Belts'' of
slightly cooler ($\simeq 4$~keV) X-ray emitting gas run around the
minor dimension of the cavity created by the radio source, while the
limb-brightened edges of the cavity are slightly hotter ($\simeq
6$~keV), perhaps as a result of heating by a bow shock driven by the
probably expanding cavity into the intracluster gas.  There is a
metallicity gradient in the X-ray emitting gas, with the highest
metallicities ($\sim$ solar) found close to the center, decreasing to
$\sim 0.3$~solar in the outer parts.  We have used the assumption of
hydrostatic equilibrium to derive a total cluster mass within 500~kpc
of $2.0 \times 10^{14} \, M_{\odot}$ and $2.8 \times 10^{14} \,
M_{\odot}$ for a constant and centrally decreasing temperature
profile, respectively.  The total mass of X-ray emitting gas within
the same radius is $1.1 \times 10^{13} \, M_{\odot}$.  Thus, the gas
fraction of the cluster within $500$~kpc is $0.055$ and $0.039$ for
the constant and centrally decreasing temperature profiles,
respectively.

\end{abstract}


\section{Introduction}

Cygnus~A is the best known and nearest example ($z=0.0562$) of a
powerful FR~II radio galaxy in a cluster of galaxies (e.g., Barthel \&
Arnaud 1996).  At X-ray wavelengths, Cygnus~A was first detected by
\citet{gia72} using the \emph{Uhuru} satellite.  \emph{Einstein}
imaging data subsequently showed the radio source to reside in
$10^{14} \, M_{\odot}$ of hot gas, extending some 1~Mpc (assuming
$H_{\rm o} = 50$~km~s$^{-1}$~Mpc$^{-1}$) to the NW of Cygnus~A
\citep{arn84}.  A dense core of hot gas, centered on the radio galaxy,
was found to have an apparent cooling flow rate of $\sim 90 \,
M_{\odot} \, \rm yr^{-1}$ \citep{arn84}.  \emph{EXOSAT} and
\emph{HEAO-1} observations yielded a temperature of $kT \simeq
4$--$5$~keV for the intracluster gas, with the radio galaxy modeled as
an absorbed power-law \citep{arn87}.  Subsequent observations with
\emph{Ginga} \citep{uen94} and \emph{ASCA} (Sambruna, Eracleous, \&
Mushotzky 1999) gave a higher temperature of $kT \simeq 7$--$8$~keV
and an iron abundance of $\sim 0.3$ times solar.  From \emph{ROSAT}
Position Sensitive Proportional Counter (PSPC) and High Resolution
Imager (HRI) observations, \citet{rf96} suggest that the gas
temperature within the inner 50~kpc is significantly cooler ($kT
\simeq 2.5$~keV) than the ambient cluster temperature at radii
$\simgreat 180$~kpc, and that the cooling flow rate is $\sim 250 \,
M_{\odot} \, \rm yr^{-1}$.  In a recent analysis of the \emph{ASCA}
data, \citet{mar98} and Markevitch, Sarazin, \& Vikhlinin (1999) have
found evidence for a merger between the two sub-clusters whose gas
temperatures are $kT \simeq 4$--$5$~keV.  The bi-modal redshift
distribution of the galaxies in Cygnus~A is also suggestive of a
merger, although the density of galaxies does not correspond well to
the distribution of X-ray emitting gas \citep{owe97}. From an analysis
of the undeconvolved \emph{ASCA} Gas Imaging Spectrometer data,
\citet{whi00} derived the overall cluster temperature and iron
abundance to be $9.49 \pm 0.23$~keV and $0.67 \pm 0.03$ solar,
respectively.  The data were also modeled with a strong cooling flow
of $\approx 500 \, M_{\odot} \, \rm yr^{-1}$, a high ambient cluster
temperature of $\approx 40$~keV, and a metallicity close to solar
values.

This paper is devoted to a \emph{Chandra} study of the intracluster
gas in the Cygnus~A cluster of galaxies.  The morphology and radial
profile of the X-ray emitting gas are discussed in
Sections~\ref{sec-morphology} and ~\ref{sec-radial}.  In
Sections~\ref{sec-spectrum} and ~\ref{sec-observed}, we derive the
spectra of the gas and the observed radial dependences of the
temperature and abundance.  Section~\ref{sec-deprojected} is devoted
to a deprojection of the observed spectral brightness distribution,
providing the radial dependences of the temperature, abundance,
emissivity, density, and pressure of the intracluster gas.  These
parameters are then used (Section~\ref{sec-mass}) to obtain the
distributions of the gas mass and total mass of the cluster.  In
Section~\ref{sec-filaments}, we discuss the extended filaments and
belts of X-ray emission around the cavity in the intracluster gas
which has been created by the radio source.  Conclusions are
summarized in Section~\ref{sec-conclusions}.

We adopt $H_{\rm o} = 50$~km~s$^{-1}$~Mpc$^{-1}$ and $q_{\rm o} = 0$
in this paper, which gives $1^{\prime\prime} = 1.51$~kpc, an angular
size distance of $d_{\rm A} = 310.5$~Mpc, and a luminosity distance to
Cygnus~A of $d_{\rm L} = 346.4$~Mpc.

\section{Observations and Data Reduction} \label{sec-observations}

Cygnus~A has been observed on three occasions with the \emph{Chandra}
ACIS (ACIS is described by G. Garmire et al., in preparation).  The
results from an analysis of the radio hot spots (Wilson, Young, \&
Shopbell 2000, hereafter Paper I) and the central nucleus (Young et
al. 2001, hereafter Paper II) are presented elsewhere.  The rationale
for the three observations is described in Paper~II.

We concern ourselves here with a study of the surrounding intracluster
gas, and analyze only data acquired during the $\simeq 35$~ks exposure
on May 21, 2000 (obsid 360), which provided an observation of Cygnus~A
over a wide field.  The data reduction and analysis were done using
\emph{Chandra Interactive Analysis of Observations (CIAO)} software
version 2.1, and the reprocessed (on January 25, 2001) event files.
We created a new level 2 events file, applying the latest gain
corrections (of January 25, 2001), the same Good Time Intervals as in
the existing level 2 events file, and \emph{ASCA} grades 0, 2, 3, 4,
and 6.  Periods of non-quiescent background (i.e., flares or data
dropouts due to telemetry saturation) were removed from the data.
This was achieved by creating a light curve of the whole S3 chip,
excluding the brightest sources of X-ray emission, and removing events
$\pm 3\sigma$ from the mean count rate.  This gives an effective
exposure time of $34.3$~ks (corrected for dead-time in the detector).

The X-ray emission from the cluster extends over the whole S3 chip, so
our analysis procedure followed that described in \citet{mar00}.
Spatial variations in the detector background were estimated from
several ACIS blank-field observations\footnote{See
http://hea-www.harvard.edu/$\sim$\,$\!$maxim/axaf/acisbg/ for further
details regarding the ACIS background fields.} with a total exposure
of $\simeq 137$~ks on the S3 chip.  These data were taken at the same
focal plane temperature as our \emph{Chandra} observation (i.e.,
$-120^{\circ}$C).  In the 10--11~keV band, where few or no cosmic
X-rays are expected, the observed count rate is $5 \pm 1$\% higher in
our observation than in the blank field observations (where the quoted
error is only the random component).  This excess count rate is
comparable with the systematic error in the background rate from field
to field \citep{mar00}.  Therefore, in the following sections, we have
used a background rate that is 5\% higher than the value obtained from
the blank field observations.  Finally, an exposure map for the S3
chip was created from knowledge of the satellite pointing direction
and the effective area across the detector.

In Section~\ref{sec-deprojected}, we present a deprojection of the
X-ray emission of the cluster gas associated with Cygnus~A.  In brief,
we derive the radial temperature, abundance, and emissivity using the
{\sc XSPEC} mixing model ``projct''\footnote{Publicly available in
{\sc XSPEC} v11.1} written by one of us (KAA).  This mixing model sums
model spectra from ellipsoidal shells to give the observed, projected
spectral emission of each elliptical annulus.  The number of
ellipsoidal shells is taken to be the same as the number of annuli,
with each shell projecting onto one annulus. In the inner regions of
the cluster, the shells are assumed to be prolate, with the major axis
in the plane of the sky.  In the outer regions, the shells are
spherical.  Thus, there are no free parameters associated with the
geometry.  The ``projct'' model allows the annuli to have different
semi-major axes, semi-minor axes, and orientations but they are
required to have the same centroid.

\section{Results}

\subsection{Morphology of The Extended X-ray Emission} \label{sec-morphology}

An unsmoothed \emph{Chandra} X-ray image of the region of the radio
source Cygnus~A was shown in Figure~1 of Paper I (see also
Figure~\ref{fig11} of present paper).  An alternative representation
can be obtained by adaptively smoothing the image (the \emph{CIAO}
program {\sc CSMOOTH}).  In this process, the width of the
two-dimensional Gaussian smoothing profile is a function of the signal
to noise ratio of the data.  Pixels with signal $> 5\sigma$ above the
local background were left unsmoothed while for the remaining pixels,
the width of the Gaussian was increased until the convolved signal at
that location exceeded the local background by $3\sigma$.  The
resulting image is shown in Figure~\ref{fig1}.

Inspection of the unsmoothed (Figure~1 of Paper I; Figure~\ref{fig11}
of present paper) and smoothed (Figure~\ref{fig1}) images reveals the
following features (see Wilson, Young, \& Shopbell 2001 for an earlier
summary).  The dominant structure is a roughly elliptical (presumably
prolate spheroidal in 3-d) feature with semi-major axis $\simeq
1.\!^{\prime}1$.  The major axis coincides well with that of the radio
source, and the radio hot spots are seen as compact X-ray sources at
the ends of the major axis (Paper I).  This structure appears to be
the observational manifestation of the intracluster gas which has been
swept up and compressed by a cavity which is expected to be inflated
in the intracluster gas by relativistic material which has passed
through the hot spots at the ends of the jets \citep{sch74}.  X-ray
emission around the cavity seems to be edge-brightened.  Curved linear
structures run along the major axis from the nucleus to each pair of
hot spots and may be the X-ray manifestations of the jets; they will
not be discussed further in this paper.  There is extended X-ray
emission associated with the nucleus (Paper II).  Bright curved bands
of X-ray emission extend along the minor axis of the ellipse.  Their
morphology, particularly in the unsmoothed image, is suggestive of
``belts'' of gas extending around the equator of the prolate
spheroidal structure.  Since the dominant radio jet is on the NW side
of the nucleus, it is usually supposed that this is the near side of
the radio source.  If the ``belts'' are circular and in the equatorial
plane of the cavity, their sense of curvature indicates that they are
predominantly on the far side.  The reason for the absence of belts
curved concave to the NW (as would be expected for the near side gas)
is unclear.  As we shall see (Section~\ref{sec-filaments}), the belts
are gas with a temperature of $\simeq 4$~keV and could represent a
large-scale accretion disk, with gas flowing in to Cygnus~A.  The
existence of the cavity and evidence for its edge-brightening was
inferred from lower resolution \emph{ROSAT} HRI observations by
\citet{car94} and a numerical simulation of its dynamics was presented
by \citet{cla97}.  A more detailed analysis of the cavity will be
presented elsewhere \citep{ws02}.

An image of the X-ray emission on chip S3 (size $8^{\prime} \times
8^{\prime}$) is shown in Figure~\ref{fig2}.  The background has been
subtracted from the raw image and the result has been corrected for
variations in vignetting and relative exposure.  The final image has
been convolved with a two-dimensional Gaussian profile of width
$\sigma = 5\, \rm pixels$ ($= 2.\!^{\prime\prime}46$).  The radio
galaxy is located towards the upper right, at the aimpoint of the S3
chip.  It is clear from this image that the spatial extent of the
X-ray emitting gas is larger than the field of view of the S3 chip, as
already known from earlier observations (e.g., Reynolds \& Fabian
1996).  In addition to the galaxy nucleus and radio hot spots, a total
of 19 compact sources were detected on the S3 chip with a significance
threshold of $10^{-6}$ using the \emph{CIAO} program {\sc
WAVDETECT}\footnote{A significance threshold of $10^{-6}$ will give
approximately one false source for the whole S3 chip (see Section
9.2.3.4 of the Detect User's Guide version 1.0 which is available at
http://cxc.harvard.edu/udocs/docs/swdocs/detect/html/MDETECT.html)}.
However, none of the galaxies listed in Table~1 of \citet{owe97}, and
visible within the field of view of the S3 chip, was detected.  We
conclude that these compact X-ray sources are most likely foreground
(e.g., stars) or background (e.g., AGN) objects.

An image of the whole \emph{Chandra} field is shown in
Figure~\ref{fig3}.  This image has been binned to a pixel size of
$3.\!^{\prime\prime}94 \times 3.\!^{\prime\prime}94$.  The bright
source near the center of the image is Cygnus~A.  The X-ray emission
from the intracluster gas extends at least $8^{\prime}$ from the radio
galaxy.  A second, fainter source of extended X-ray emission is seen
some $12^{\prime}$ to the NW of Cygnus~A, and is partially obscured by
the boundary dividing the S1 and S2 chips.  This second source was
detected by \emph{ROSAT} \citep{rf96} and by \emph{ASCA}
\citep{mar98,mar99}.  It is natural to presume that this second
extended source represents another cluster of galaxies (see Markevitch
et al. 1999).  There is also evidence for extended X-ray emission
between the two clusters.

\subsection{Radial Profile of the Intracluster Gas} \label{sec-radial}

A radial profile of the X-ray surface brightness was determined by
extracting, from the S3 chip, counts in the 0.75--8~keV band in
circular annuli of width 5~pixels ($= 2.\!^{\prime\prime}46$) and
centered on the radio galaxy.  We have ignored deviations in the
symmetry due to the prolate morphology of the intracluster gas in the
inner regions of the cluster.  A similar approach to that used in
making Figure~\ref{fig2} was adopted for removal of background and
correction for telescope vignetting.  For regions of the chip within
$1.\!^{\prime}23$ of the radio galaxy, the annuli were restricted to
two pie-shaped regions some $50^{\circ}$ and $70^{\circ}$ in angular
extent, chosen to avoid emission near the radio lobes and hot spots.
In the outer regions of the chip, compact sources of X-ray emission
were excluded from the radial profile.  The final profile has been
binned so that the signal to noise ratio in each bin exceeds $25$
(Figure~\ref{fig4}).

The radial surface brightness profile, $S(\theta)$, has been modeled
by an isothermal $\beta$ profile \citep{sb77}:
\begin{eqnarray}
S(\theta) = S_{\rm o} \times \left[ 1 + (\theta/\theta_{\rm c})^{2}
\right]^{-3\beta + 0.5}, \label{eqn-king}
\end{eqnarray}
where $S_{\rm o}$, $\theta$, $\theta_{\rm c}$, and $\beta$ are the
central surface brightness, angular distance from the center, the
angular core radius, and the slope parameter, respectively.  In
fitting the model, we have used only data at radii
$6^{\prime\prime}$--$16^{\prime\prime}$ and $\geq 37^{\prime\prime}$,
chosen to avoid, as much as possible, the increase in surface
brightness around the edge of the cavity and at the nucleus of the
radio galaxy.  The model provides a poor description of the profile
even when these anomalous regions are excluded, with $\chi^{2} =
111.5$ for 51 degrees of freedom (d.o.f.); the best-fit parameters are
$\theta_{\rm c} \simeq 18$~arcsec and $\beta \simeq 0.51$.  We note
that the isothermal $\beta$ profile is flat inside $\theta_{\rm c}$,
and thus small values of $\theta_{\rm c}$ may indicate that there is
emission in excess of that expected from the ambient intracluster gas.
\citet{car94} determined $\theta_{\rm c} = 35 \pm 5$~arcsec and $\beta
= 0.75 \pm 0.25$, using the \emph{ROSAT} HRI data at radii $\geq
6^{\prime\prime}$.  If we restrict our analysis to the data within
radii $6^{\prime\prime}$--$90^{\prime\prime}$, the same radial extent
as that used in \citet{car94}, then we find $\theta_{\rm c} \simeq
29^{\prime\prime}$ and $\beta \simeq 0.67$.  An isothermal $\beta$
profile with these parameters, when extrapolated to larger radii,
underestimates the X-ray surface brightness.

The data have also been modeled with the analytic expression of Suto,
Sasaki, \& Makino (1998) for the Navarro et al. (1996, 1997, hereafter
NFW) profile:
\begin{eqnarray}
S(\theta) = S_{\rm o} \times \left[ 1 + (\theta/\theta_{\rm c})^{\xi}
\right]^{-\eta}, \label{eqn-nfw}
\end{eqnarray}
where $\xi = -4.99 + 7.055 \times B^{-1/30}$ and $\eta = -0.68 +
0.2226 \times B^{1.1}$.  The parameter $B$ is $(4 \pi G \mu m_{\rm p}
\delta_{\rm c} \rho_{\rm c 0} r_{\rm s}^{2})/kT$, where $\delta_{\rm
c}$ is the characteristic density of the cluster, $\rho_{\rm c 0}$ is
the critical density of the universe at redshift $z = 0$, $r_{\rm s}$
is the characteristic scale radius of the cluster, $\mu$ is the mean
molecular weight ($\mu = 0.6$ if no metals present), $m_{\rm p}$ is
the mass of the proton, $k$ is Boltzmann's constant, $T$ is the
temperature of the gas, and $G$ is the gravitational constant
\citep{sut98}.  This model also provides a poor description of the
profile (again, the anomalous regions were omitted) with $\chi^{2} =
145.6$ for 51 d.o.f.; the best-fit parameters are $\theta_{\rm c}
\simeq 17$~arcsec and $B \simeq 7.4$.  The NFW profile may not be an
accurate description of the density of the dark matter halo within the
inner regions of the cluster, due to the poor spatial resolution of
the $N$-body simulations \citep{fm97,moo98}.  However, we have also
tried to model the data with a generalized form of the profile, in
which $\xi$ and $\eta$ are free parameters, without success.  It is
possible that the central $30^{\prime\prime}$ region of the cluster
are not in hydrostatic equilibrium (there a powerful radio source
located at the center of the cluster) or that the X-ray surface
brightness increases towards the center of the cluster as a result of
a cooling flow, which would explain why the data are not described by
either the isothermal $\beta$ or NFW profiles.  Unfortunately,
modeling the data at radii $\geq 37^{\prime\prime}$ does not allow a
determination of the core radius for either description of the
profile.  Instead, a power-law of $S(\theta) \propto \theta^{-1.993
\pm 0.016}$ (errors here and elsewhere in this paper are $90$\%
confidence for one interesting parameter, $\Delta\chi^{2} = 2.706$)
provides a reasonable description of the profile with $\chi^{2} =
66.3$ for 48 d.o.f.

\subsection{Spectra of the Intracluster Gas} \label{sec-spectrum}

The spectral variations in the intracluster gas are illustrated in
Figure~\ref{fig5}, which is a color map of the softness ratio
(1--2~keV/2--8~keV) for the same region as shown in Figure~\ref{fig2}.
We have extracted images in the 1--2~keV and 2--8~keV bands, as input
to the program {\sc ADAPTIVEBIN} version 0.1.2 \citep{sf01}.  The
resulting color image was adaptively binned so that the maximum
fractional error in the color was 0.1 at each location (see color bar
on right of Figure~\ref{fig5}).  Background was subtracted from the
images, so that the colors represent the softness ratios of the
celestial X-ray emission.  The color image shows that emission inside
the central $\simeq 1^{\prime}$ is softer than that further out.  The
hardest emission is located at the position of the nucleus, which is
heavily absorbed (Arnaud et al. 1987; Ueno et al. 1994; Sambruna et
al. 1999; Paper II).

Our goal in this paper is to investigate the spatial variations of the
properties (e.g., density, temperature, and abundances) of the
intracluster gas in the inner region of the Cygnus~A cluster (i.e.,
that which falls on chip S3 in our observation).  We began by
extracting spectra from several annuli centered on the radio galaxy
(see Figure~\ref{fig6}) using PI channels, which account for
variations in the gain across each of the extraction regions.  We
created response matrices appropriate for our spectra by weighting the
position dependent response and effective area by the number of counts
in the corresponding region.  Care was taken to avoid emission powered
by the radio galaxy itself; this is the reason only sectors of
elliptical annuli (Figure~\ref{fig6}) were used in the inner regions
(see Table~\ref{tbl-1} for the parameters of the ellipses used for
each annulus).  The angular extents of the sectors used for the outer
annuli were limited by the size of the CCD chip (Figure~\ref{fig2}).

Initially, the spectrum of \emph{all} the intracluster gas within the
S3 chip, excluding the contribution from the radio galaxy and other
compact sources of X-ray emission, was modeled by an absorbed, single
temperature mekal \citep{mew95}, using {\sc XSPEC} version 11.0.1
\citep{arn96}.  The absorption cross-sections of \citet{mm83} and
abundances of \citet{ag89} were used throughout.  This model (model 1)
gives a poor fit to the data with $\chi^{2} = 736.2$ for 489
d.o.f. (see Table~\ref{tbl-2}).  The best-fit temperature,
metallicity, and equivalent hydrogen absorbing column density are $kT
\simeq 8.3$~keV, $Z \simeq 0.35 \, Z_{\odot}$, and $N_{\rm H} \simeq
2.8 \times 10^{21}$~cm$^{-2}$, respectively.  The combined spectrum,
together with the residuals to this model fit, are shown in
Figure~\ref{fig7}.  There are large residuals near the M-edge of
iridium, which arises from X-ray absorption in the telescope mirror
surface.  Large residuals near the iridium M-edge were also found in
the spectra of two high temperature clusters, A665 and A2163, observed
with the front-illuminated CCDs on \emph{Chandra} \citep{mv01}.
Ignoring the data between 1.8 and 2.2~keV (as Markevitch \& Vikhlinin
have done) gives a better, although still unacceptable, fit of
$\chi^{2} = 548.2$ for 461 d.o.f., with the observed temperature,
metallicity, and absorbing column density essentially unchanged from
their former values.  For comparison, the Medium Energy experiment on
\emph{EXOSAT} measured a temperature of $4.1^{+5.6}_{-1.5}$~keV and an
iron abundance of $0.59^{+0.68}_{-0.35}$ times solar for the
intracluster gas \citep{arn87}.  Similarly, a temperature of
$7.3^{+1.8}_{-1.3}$~keV and an iron abundance of $\simeq 0.32$ times
solar were found from spectral fits to the \emph{Ginga} Large Area
Counter data \citep{uen94}.  Both of these experiments used
non-imaging proportional counters, which included in their field of
view emission from both the radio galaxy and the intracluster gas
outside chip S3.  More recent observations with \emph{ASCA}, which
also refer to a larger region than ours, measured the overall cluster
temperature and abundance to be $kT \simeq 8$~keV and $\sim 0.6$ times
solar, respectively \citep{mar98,mar99,sam99,whi00}.

There is evidence from \emph{ASCA} observations of clusters for an
Si/Fe abundance ratio which increases with increasing gas temperature
\citep{fuk98}.  An increase in the Si/Fe abundance would produce more
line emission at $1.84$--$1.86$ (Si~{\sc xiii}), $2.01$
(Si~Ly$\alpha$), and $2.38$ (Si~Ly$\beta$)~keV, which may contribute
to the positive residuals near the iridium M-edge.  Thus, we have
replaced the mekal model with a vmekal model, in which H and He have
solar abundances, C and N have $0.3$ solar abundances, and the
remaining metals are permitted to have variable abundances.  This
model provides a slightly better description of the data than model~1
with $\chi^{2} = 726.6$ for 479 d.o.f.  The best-fit temperature, Si
abundance, Fe abundance, and equivalent hydrogen absorbing column
density are $kT \simeq 8.3$~keV, $Z (\rm Si) \simeq 0.38 \,
Z_{\odot}$, $Z (\rm Fe) \simeq 0.33 \, Z_{\odot}$, and $N_{\rm H} =
2.8 \times 10^{21}$~cm$^{-2}$, respectively.  The Si/Fe abundance
ratio is $\simeq 1.1$, which is smaller than that observed in other
clusters with a similar gas temperature \citep{mus96,fuk98}.  Thus, an
increased Si/Fe abundance ratio is not required by the data, but the
inaccuracies in the response matrix at the energy of the Si line
emission makes an accurate determination of the Si/Fe abundance ratio
difficult.  For the remainder of this paper, we shall assume a Si/Fe
abundance ratio of unity.

The Galactic column density towards Cygnus~A is $N_{\rm H} (\rm Gal) =
3.5 \times 10^{21}$ cm$^{-2}$ \citep{dl90}, which is higher than that
inferred from model~1.  One possibility is that the Galactic column
density is lower than that estimated from Dickey \& Lockman's
observations.  Another explanation for this discrepancy is that the
soft X-ray response of the detector is lower than assumed.  However,
we have used only data above 0.75~keV, where the ACIS S3 chip is well
calibrated.  Additionally, when the radial dependence of the column
density was first derived (see Section~\ref{sec-observed}), we found
that the best-fit column density decreased with increasing distance
from the radio galaxy, reaching $N_{\rm H} \simeq 1.8 \times
10^{21}$~cm$^{-2}$ at the outer annulus.  This effect could represent
a real variation in the column density.  Alternatively, given the low
Galactic latitude of Cygnus~A ($b = 5.\!^{\circ}76$), this apparent
spatial dependence of $N_{\rm H}$ could result from the presence of
diffuse, soft X-ray emission from our Galaxy.  To test this idea, we
have added to model~1 a second mekal component, with solar abundances,
to represent this possible Galactic emission, with both mekals
constrained to be absorbed by the same column density (model~2).  The
improvement in the fit compared with that for model~1 is significant
at $> 99$\% confidence (on the basis of an \emph{F} test for two
additional free parameters), with $\chi^{2} = 709.4$ for 487 d.o.f.;
the best-fit parameters are given in Table~\ref{tbl-2}.  The observed
0.75--8~keV flux of the high temperature, cluster component is $2.7
\times 10^{-11}$ erg cm$^{-2}$ s$^{-1}$.  There is no improvement (at
$>90$\% confidence) in the fit when we allow the abundance of the low
temperature, Galactic thermal component to vary.  The observed surface
brightness of this possible Galactic thermal component is $5.2 \times
10^{-15}$ erg s$^{-1}$ cm$^{-2}$ arcmin$^{-2}$ in the 0.75--8~keV
band.

\subsection{Observed radial dependence of temperature, abundance, 
and emissivity} \label{sec-observed}

The next step was to model the individual annuli with an absorbed
mekal.  We assumed there is little variation in the Galactic column
over the cluster, and so assumed the same column for all annuli.  Two
mekal components were used for each annulus.  The first was assumed to
have uniform temperature, surface brightness, and abundance (fixed at
solar), and was taken to represent the possible foreground Galactic
emission (see Section~\ref{sec-spectrum}).  The second, intended to
represent the cluster, permitted variable temperature, abundance, and
brightness, to derive the radial variations of these properties in the
cluster.  The results from this model, in which all annuli were fitted
simultaneously, are given in Table~\ref{tbl-3}.  We note that
variations in the absorption column density are not required by the
data for this model.  The inferred intracluster gas temperature varies
from $\sim 8$~keV more than $\sim 2^{\prime}$ from the center to
$\simeq 5$~keV some $30^{\prime\prime}$ from the center, with the
coolest gas immediately adjacent to the cavity created by the radio
galaxy (Figure~8a).  This temperature variation is compatible with the
softness ratio map in Figure~\ref{fig5}.  There is also clear evidence
for a metallicity gradient in the X-ray emitting gas, with the highest
metallicities ($\sim 0.6$--$0.8$ solar) found in the inner annuli,
decreasing to $\sim 0.3$ solar in the outer parts (Figure~8b).

It is interesting to compare these results with those obtained
previously.  From the \emph{ROSAT} PSPC data, \citet{rf96} find
evidence for $2.5^{+0.7}_{-0.4}$~keV gas within $1^{\prime}$ of the
radio nucleus.  Their temperature is lower than that determined from
our modeling of the individual annuli.  However, we note that there is
evidence for gas at a temperature of $\simeq 4$~keV in the ``belts''
around the cavity (see Section~\ref{sec-filaments} below), which is
marginally consistent with their upper limit; in fact, the PSPC
spectrum within $1^{\prime}$ of the nucleus may be dominated by the
emission from this 4~keV gas.  The temperature derived from the PSPC
data at radii $\geq 100$~kpc is consistent with the values determined
from the \emph{Chandra} data.  Using \emph{ASCA} data, \citet{mar99}
have constructed a map of the projected gas temperature within the
Cygnus~A cluster.  However, their map has poor spatial resolution,
with most of the intracluster gas within the S3 chip residing in
region~1.  \citet{mar99} measured a gas temperature in the range
$\simeq 5$ to $\simeq 7$~keV for X-ray emission within this region,
which is consistent with the temperature of the integrated
intracluster gas on S3 (Section~\ref{sec-spectrum}). \citet{mar99}
identified regions of potentially higher and lower gas temperatures,
but most of this emission resides on the other \emph{Chandra} CCDs.
\citet{whi00} has performed a re-analysis of the \emph{ASCA} data, and
finds that the gas has a temperature in the range $\simeq 6$ to
$\simeq 8$~keV in the region covered by the S3 chip, and a metallicity
of $\simeq 0.2$--$0.3$ solar.

\subsection{Deprojected radial dependence of temperature, abundance, 
and emissivity} \label{sec-deprojected}

The X-rays observed from each annulus contain emission from all gas
along our line of sight in that direction.  Thus, to obtain the
physical parameters of the thermal gas as a function of distance from
the center, it is necessary to deproject the observed spectral
brightness distribution.  In performing this deprojection, we have
assumed that the X-ray emission from the cluster around the radio
galaxy is distributed in a series of prolate ellipsoidal shells, whose
major axis is along the projected radio jet axis at a position angle
(P.A.) of $\simeq 116.\!^{\circ}3$ and is in the plane of the
sky. Guided by the X-ray isophotes (Figure~\ref{fig2}), we assumed
that the ellipticity of these shells steadily decreases until the
shells become spherical some $2.\!^{\prime}5$ from the center.  We
feel this geometry is likely to be a good approximation to that of the
intracluster gas in this inner part of the cluster (there is an
extension $\sim 12^{\prime}$ to the NW, but this region is not on chip
S3).  As a narrow line radio galaxy, the radio jet and lobes of
Cygnus~A are expected to be close to the plane of the sky, a geometry
favored by \citet{hr74} on independent grounds.  \citet{car96} state
that the average ejection angle could be $\phi > 60^{\circ}$ from our
line of sight, while \citet{sor96} favor $55^{\circ} < \phi <
80^{\circ}$ based on VLBI observations.  Thus our assumption that the
major axis of the the prolate spheroid lies in the plane of the sky
may not be too far wrong.  One limitation of our analysis is that our
outer annulus (Figure~\ref{fig6}) does not represent the true edge of
the intracluster emission (cf. Reynolds \& Fabian 1996), so there will
be faint, more extended intracluster emission projected onto it.

As described in Section~\ref{sec-observations}, we deprojected the
observed emission using the {\sc XSPEC} model ``projct''.  Each shell
was modeled as a mekal (thermal) plasma.  The spectral analysis was
performed in the following way.  First, a uniform brightness mekal
model (representing the Galactic emission) was added to a mekal model
for the outer shell and the projected result was compared with the
observed spectrum of the outer annulus (annulus 12).  Both emission
components were assumed to be absorbed by the same column and the
model parameters for absorption and Galactic emission were
subsequently fixed at their respective best-fit values given in
Table~\ref{tbl-3}.  In this way, we derived the temperature,
abundance, and emissivity for the outer shell of intracluster gas.
Second, the spectra from shells 11 and 12 were modeled as above.  The
temperature, abundance, and emissivity for shell 12 were fixed at the
values already obtained, allowing us to derive the deprojected
temperature, abundance, and emissivity for shell 11.  This procedure
was repeated for the remaining shells, and the results for all shells
are given in Table~\ref{tbl-4}.  An alternative approach would be to
sequentially model the various annuli, but to allow their parameters
to be free in the final fitting in which all the projected mekal
models are compared with the data.  This approach has the disadvantage
that the final modeling is insensitive to the fainter outer regions,
for which the derived parameters are thus unreliable.  For this
reason, we consider our approach, in which one annulus is added at
each modeling and then ``frozen'' in subsequent iterations, to be
superior.

The properties of the intracluster gas for each of the spheroidal or
spherical shells are given in Table~\ref{tbl-5}.  The properties of
the gas in the outer shell (shell 12) are not listed since the volume
of this shell is smaller than the volume of the intracluster gas
projected onto it, rendering the calculated properties unreliable.
The pressure and cooling time of the gas are $n_{\rm e} k T$ and
$n_{\rm e} k T V/L_{\rm X}$, respectively.  We have used the usual
notation here, in which $n_{\rm e}$ is the electron density, $V$ is
the volume of the shell, and $L_{\rm X}$ is the 2--10~keV rest-frame
luminosity after correcting for absorption. In Figure~\ref{fig9}, we
show the radial dependence of deprojected temperature, electron
density, abundance, and gas pressure.  The peak in gas pressure
between $90$ and $100$~kpc is not significant given the errors on the
temperature at this location.  The deprojected and observed
temperatures and abundances are consistent with each other, which
suggests that the emission observed in the inner regions of the
cluster dominates the fainter outer emission projected onto it.  It is
noteable that the cooling time is less than the Hubble time for radii
$\simless 200$~kpc.

\subsection{Distribution of Mass in the Cluster} \label{sec-mass}

Under the assumption of hydrostatic equilibrium and spherical symmetry
of the inter-cluster gas, the mass of the cluster, $M$, within a
radius, $r$, is
\begin{eqnarray}
M(<r) = -\frac{k T(r) r^{2}}{\mu m_{\rm p} G} \left(\frac{1}{n_{\rm
e}}\frac{d n_{\rm e}}{d r} + \frac{1}{T}\frac{d T}{d r} \right)
\label{eqn-mass}
\end{eqnarray}
(e.g., Fabricant, Lecar, \& Gorenstein 1980), where $T(r)$ is the
temperature at radius $r$.  We have considered two alternative
descriptions of the temperature: (i) a constant temperature of
5.73~keV; (ii) a centrally decreasing profile of the form
\begin{eqnarray}
T(\rm keV) = a - b \, \exp(-r(kpc)/c), \label{eqn-temp}
\end{eqnarray}
where $a = 7.81$~keV, $b = 7.44$~keV, and $c = 76.4$~kpc are constants
determined from modeling the deprojected temperatures.  Hence, for the
centrally decreasing temperature profile, the second term in
parentheses in Equation~\ref{eqn-mass} may be written
\begin{eqnarray}
\frac{1}{T}\frac{d T}{d r} = \frac{1}{T} \frac{b}{c}
\exp(-r/c). \label{eqn-deriv_temp}
\end{eqnarray}
The radial dependence of the electron density has been modeled as a
broken power-law of the form
\begin{eqnarray}
n_{\rm e} (\rm cm^{-3}) = \left\{ 
\begin{array}{r@{\quad:\quad}l} 
w \, (r(\rm kpc)/x)^{-y} & r > c \\
w \, (r(\rm kpc)/x)^{-z} & \rm otherwise,
\end{array} \right. \label{eqn-density}
\end{eqnarray}
where $w = 8.24 \times 10^{-3}$~cm$^{-3}$, $x = 97.2$~kpc, $y = 1.50$,
and $z = 4.11$ are the best-fit values.  The slope of the density
profile at large radii is consistent with an isothermal $\beta$
profile of $\beta = 0.5$ (the electron density is $n_{\rm e} \propto
r^{-3\beta}$ for $r >> \theta_{\rm c}$).  A similar value for $\beta$
was derived directly from the X-ray surface brightness profile in
Section~\ref{sec-radial}.  The first term in parentheses in
Equation~\ref{eqn-mass} may be written
\begin{eqnarray}
\frac{1}{n_{\rm e}}\frac{d n_{\rm e}}{d r} = \left\{
\begin{array}{r@{\quad:\quad}l}
-(y/x) \, (r/x)^{-1} & r > c \\ 
-(z/x) \, (r/x)^{-1} & \rm otherwise.
\end{array} \right. \label{eqn-deriv_density}
\end{eqnarray}

Substituting Equations~\ref{eqn-deriv_temp} and
\ref{eqn-deriv_density} into Equation~\ref{eqn-mass} gives the
integrated mass of the cluster, including the contributions from dark
matter, intracluster gas, and individual galaxies.  The radial
dependence of enclosed mass $M(<r)$ is shown in Figure~\ref{fig10} for
radii 80--500~kpc.  We have also calculated the integrated mass of the
intracluster gas, $M_{\rm gas}$, (also shown in Figure~\ref{fig10})
for the same range of radii, from
\begin{eqnarray}
M_{\rm gas}(<r) = \mu \, m_{\rm p} \int n_{\rm e}(r) \, {\rm d}V,
\label{eqn-gas_mass}
\end{eqnarray}
where ${\rm d}V$ is the volume of the shell containing gas at density
$n_{\rm e}(r)$ (which is given by Equation~\ref{eqn-density}).  The
integrated mass of gas within the inner 80~kpc of the cluster was
determined by assuming a constant density of $n_{\rm e} = w (80~\rm
kpc)/x)^{-z} = 0.018$~cm$^{-3}$.  From the above analysis, we find the
total mass of the cluster, within 500~kpc, is $2.0 \times 10^{14} \,
M_{\odot}$ and $2.8 \times 10^{14} \, M_{\odot}$ for the constant and
centrally decreasing temperature profiles, respectively.  This
compares favorably with the cluster mass of $10^{14} \, M_{\odot}$,
derived from the \emph{Einstein} HRI data \citep{arn84}.  Within
500~kpc, the total mass of the intracluster gas is $1.1 \times
10^{13} \, M_{\odot}$.

\subsection{Extended Filaments and Belts of X-ray Emission near the 
Radio Source Cavity} 
\label{sec-filaments}

Spectra were extracted from several regions of bright, extended X-ray
emission near the nucleus (see Figure~\ref{fig11}). The regions were
chosen so that they included the curved belts of X-ray emission which
appear to encircle the radio source cavity, as well as the bright
bands of emission around the edge of the cavity.  Care was taken to
avoid the curved linear structures which extend from the nucleus to
radio hot spots as these regions are apparently related to the radio
activity which is outside the scope of this paper.  Prior to modeling,
the spectra were re-binned so that there were at least 20~cts
bin$^{-1}$ and the analysis was conducted over the full 0.5--10 keV
band, as the diffuse, X-ray emission from our Galaxy is insignificant
compared to this emission from these bright inner regions of the
intracluster medium.

We have modeled each spectrum with a single temperature mekal,
absorbed by a column of cold gas, which provides an acceptable fit to
all 10 spectra (see table~\ref{tbl-6} for details).  The gas
associated with the regions of the curved belts (C, D, E, F, G, and I)
has a temperature of $\simeq 4$~keV, which is cooler than that
observed some $30^{\prime\prime}$ from the center of the cluster (see
Section~\ref{sec-observed} and Table~\ref{tbl-3}).  Gas located in the
bands of X-ray emission around the edge of the cavity (regions A, B,
H, and J) has a higher temperature of $\simeq 6$~keV, which is
slightly higher than that inferred ($\simeq 5$~keV) in the adjacent,
inner regions of the intracluster gas (Section~\ref{sec-deprojected}
and Tables~\ref{tbl-3} and \ref{tbl-4}).  This slightly higher
temperature might result through heating of the gas by the probably
expanding cavity, by means of the bow shock driven into the
intracluster gas.  The metallicities ($\sim 0.8$--$1$ solar) are
consistent with, if somewhat higher than, those observed some
$30^{\prime\prime}$ from the center.  There appears to be a slight
gradient in the absorption column density, with the higher columns
observed north of the nucleus.  We note that there is no obvious
relationship between measured column density and size of the
corresponding extraction region, which would result if diffuse, X-ray
emission from our Galaxy was significant.  However, the uncertainties
on the absorption column densities are comparable to the range of
columns, so it is unclear whether this gradient is real.

\section{Conclusions} \label{sec-conclusions}

We have analyzed the properties of the intracluster gas around
Cygnus~A using an observation with the ACIS-S aboard \emph{Chandra}.
The prolate morphology of the X-ray emission from the intracluster gas
within $\simeq 2^{\prime}$ of the radio galaxy appears to be the
result of gas being swept up by a cavity created in the intracluster
gas by the radio source.  At larger radii, the X-ray emission from the
intracluster gas appears to be roughly spherical, and extends at least
$8^{\prime}$ from the radio galaxy.  A second, fainter source of
extended X-ray emission is seen some $12^{\prime}$ to the NW of
Cygnus~A.  The radial X-ray surface brightness profile, $S(\theta)$,
cannot be modeled by either an isothermal $\beta$ model \citep{sb77}
or the analytic expression of \citet{sut98} for the NFW
\citep{nav96,nav97} profile.  However, we find that the surface
brightness at radii $\geq 37^{\prime\prime}$ is adequately described
by a power-law $S(\theta) \propto \theta^{-1.993 \pm 0.016}$.  The
spectrum of the intracluster gas imaged on the S3 chip has been
modeled by a mekal absorbed by a single column of cold gas.  The
temperature and unabsorbed 2--10~keV rest-frame luminosity of the
intracluster gas imaged on S3 are $7.7$~keV and $3.5 \times 10^{44}$
erg s$^{-1}$, respectively.  The equivalent hydrogen column density is
$3.39 \times 10^{21}$ cm$^{-2}$, close to the Galactic value, and the
metallicity is $0.34$ times solar.  Although the data do not require
an enhanced Si/Fe abundance ratio relative to solar, as detected in
other clusters with similar gas temperatures \citep{fuk98}, the actual
Si abundance is uncertain as a result of the inaccuracies in the
response matrix at energies close to the iridium M-edge.  We also
suspect a Galactic thermal component with temperature $0.67$~keV and
observed surface brightness $\simeq 5.2 \times 10^{-15}$ erg s$^{-1}$
cm$^{-2}$ arcmin$^{-2}$ in the 0.75--8~keV band.

We have deprojected the X-ray spectra taken from several elliptical
and circular annuli centered on Cygnus~A.  The temperature of the
X-ray emitting gas drops from $\simeq 8$~keV more than $100$~kpc from
the center to $\simeq 5$~keV some $80$~kpc from the center, with the
coolest gas immediately adjacent to the cavity.  The bright ``belts''
of X-ray emission that appear to encircle the radio source cavity are
somewhat cooler ($\simeq 4$~keV), while the ``filaments'' responsible
for the limb-brightening of the cavity are slightly hotter ($\sim
6$~keV), perhaps as a result of heating by a bow shock driven by the
probably expanding cavity into the intracluster gas.  A metallicity
gradient is detected in the X-ray emitting gas, with the highest
metallicities ($\sim$~solar) found close to the center decreasing to
$\sim 0.3$~solar in the outer parts.  We have used the assumption of
hydrostatic equilibrium to derive the total gravitational mass of the
cluster as a function of radius.  The total mass of the cluster within
500~kpc is $2.0 \times 10^{14} \, M_{\odot}$ and $2.8 \times 10^{14}
\, M_{\odot}$ for a constant and centrally decreasing temperature
profile, respectively. The total mass of X-ray emitting gas within the
same radius is $1.1 \times 10^{13} \, M_{\odot}$.  Thus, the gas
fraction of the cluster within 500~kpc is $0.055$ and $0.039$ for the
constant and centrally decreasing temperature profile, respectively.

\acknowledgments

This research was supported by NASA grant NAG~81027 and by a
fellowship to ASW from the Graduate School of the University of
Maryland.  YT is supported by the Japan Society for the Promotion of
Science Postdoctoral Fellowship for Young Scientists.


\clearpage



\figcaption[f1.eps]{An image of the central region of the Cygnus~A
field in the 0.75--8~keV band.  The image has been ``adaptively
smoothed'' with a 2-d Gaussian profile of varying width (see
Section~\ref{sec-morphology} for details).  The shade is proportional
to the square root of the intensity.  The shading ranges from $0$
(white) cts pixel$^{-1}$ to $15$ (black) cts pixel$^{-1}$ (1 pixel is
$0.\!^{\prime\prime}492$ square).  Filamentary features close to the
nucleus, located at the center, are enhanced in this image. 
\label{fig1}}

\figcaption[f2.eps]{The distribution of intracluster gas in an $\simeq
8^{\prime} \times 8^{\prime}$ region (i.e., chip S3) of the Cygnus~A
field in the 0.75--8 keV band.  The background has been subtracted and
the image convolved with a 2-d Gaussian profile of width $\sigma = 5
\, \rm pixels$ ($= 2.\!^{\prime\prime}46$).  The vertical bar
represents the relation between grey scale and cts pixel$^{-1}$
s$^{-1}$ cm$^{-2}$ (i.e. count rate corrected for variations in the
effective area over the chip) in the convolved image. \label{fig2}}

\figcaption[f3.eps]{An image of the whole \emph{Chandra} field in the
0.75--8~keV band.  The image has been binned to a pixel size of
$3.\!^{\prime\prime}94 \times 3.\!^{\prime\prime}94$ and the parallel
streaks on chip S4 resulting from read-out noise have been removed
using the \emph{CIAO} program {\sc DESTREAK}.  The vertical bar
indicates the conversion from color to cts pixel$^{-1}$. \label{fig3}}

\figcaption[f4.eps]{The radial profile of the intracluster gas in the
0.75--8~keV band.  The data are shown (dots) together with the modeled
isothermal $\beta$ profile (solid line, see Section~\ref{sec-radial}).
\label{fig4}}

\figcaption[f5.eps]{A color representation of the softness ratio
(i.e., 1--2~keV/2--8~keV) in the region covered by the S3 chip
superposed on X-ray contours of the background-subtracted image in the
0.75--8~keV band (see Section~\ref{sec-spectrum}).  The X-ray image
has been adaptively smoothed with a 2-d Gaussian profile of varying
width.  The vertical bar indicates the relation between color and
softness ratio, and the pixels were rebinned so that the fractional
error in the ratio did not exceed 0.1.  The contours indicate $2.5^{n}
\times 10^{-2}$ cts pixel$^{-1}$, where $n = 1$, 2, 3, 4, 5, 6, 7, and
8. \label{fig5}}

\figcaption[f6.eps]{A background-subtracted image of the Cygnus~A
field covered by the S3 chip in the 0.75--8~keV band.  The image has
been convolved with a 2-d Gaussian profile of width $\sigma = 5 \, \rm
pixels$ ($= 2.\!^{\prime\prime}46$).  The solid black lines mark
regions from which spectra of the intracluster gas were extracted (see
Table~\ref{tbl-1}).  The numbers of the annuli are indicated.  The
vertical bar represents the relation between grey scale and cts
pixel$^{-1}$.  \label{fig6}}

\figcaption[f7.eps]{The spectrum of all the intracluster gas imaged on
chip S3 (see Section~\ref{sec-spectrum}).  The upper panel shows the
data together with the folded model (model 1, solid line, see
Table~\ref{tbl-2}).  The residuals, in units of $\sigma$, from the
best-fit absorbed mekal model are shown in the lower panel.  The data
have been binned so that the signal to noise ratio in each bin exceeds
10, but with no more than 10 spectral channels added together.
\label{fig7}}

\figcaption[f8.eps]{The observed properties of the intracluster gas in
individual annuli (see Table~\ref{tbl-1} for the parameters of the
ellipses or circles used for each annulus) surrounding the radio
galaxy.  For the elliptical annuli, the projected radius is the
average of the semi-major and semi-minor axes of that annuli.  (a) Gas
temperature.  (b) Metal abundance. \label{fig8}}

\figcaption[f9.eps]{The deprojected properties of the intracluster
gas in individual shells (see Table~\ref{tbl-1} for the parameters of
the ellipsoids or spheres used for each shell) surrounding the radio
galaxy.  For each ellipsoidal shell, the radius is the average of the
semi-major and semi-minor axis.  Derived properties are not given for
the outer shell for the reason given in Section~\ref{sec-deprojected}.
(a) Gas temperature (crosses) and electron density (solid line).  (b)
Metal Abundance (crosses) and thermal gas pressure (solid line).
\label{fig9}}

\figcaption[f10.eps]{The integrated mass profiles $M(<r)$ of the
cluster (solid and dashed lines), assuming hydrostatic equilibrium and
spherical symmetry, and the intracluster gas (dotted line) for radii
between 80 and 500~kpc.  The solid and dashed lines are for the
centrally decreasing temperature profile, given by
Equation~\ref{eqn-temp} in Section~\ref{sec-mass}, and the constant
temperature profile, respectively.  \label{fig10}}

\figcaption[f11.eps]{An unsmoothed image of the central region of the
Cygnus~A field in the 0.75--8~keV band.  The shade is proportional to
the square root of the intensity.  The shading ranges from $0$ (white)
cts pixel$^{-1}$ to $15$ (black) cts pixel$^{-1}$.  Solid white lines
mark regions from which spectra of extended X-ray emission were
extracted (see Section~\ref{sec-filaments}). \label{fig11}}

\clearpage

\begin{deluxetable}{cccccc}
\footnotesize \tablecaption{The boundaries of the individual annuli
used for spectral analysis\tablenotemark{a}. \label{tbl-1}}
\tablewidth{0pt} 
\tablehead{ 
\colhead{Annulus} & 
\colhead{$a$\tablenotemark{b}} & 
\colhead{$b$\tablenotemark{c}} &
\colhead{P.A.$_{\rm maj}$\tablenotemark{d}} & 
\colhead{P.A.$_{\rm begin}$\tablenotemark{e}} & 
\colhead{P.A.$_{\rm end}$\tablenotemark{f}} \\ 
\colhead{} & 
\colhead{(arcsec)} &
\colhead{(arcsec)} & 
\colhead{(degrees)} & 
\colhead{(degrees)} &
\colhead{(degrees)}} 
\startdata - & 61.5 & 28.3 & 116.3 & ... & ... \\ 
1 & 64.0 & 36.9 & 116.3 & (0.0, 170.0) & (70.0, 220.0) \\ 
2 & 66.4 & 44.3 & 116.3 & (0.0, 170.0) & (70.0, 220.0) \\ 
3 & 68.9 & 51.7 & 111.3 & (0.0, 170.0) & (70.0, 220.0) \\ 
4 & 71.3 & 61.5 & 111.3 & (0.0, 170.0) & (70.0, 220.0) \\ 
5 & 73.8 & 73.8 & 0 & (0.0, 170.0) & (70.0, 220.0) \\ 
6 & 86.1 & 86.1 & 0 & 345.0 & 237.0 \\ 
7 & 110.7 & 110.7 & 0 & 20.0 & 270.0 \\ 
8 & 147.6 & 147.6 & 0 & 20.0 & 270.0 \\ 
9 & 196.8 & 196.8 & 0 & 45.0 & 220.0 \\ 
10 & 246.0 & 246.0 & 0 & 90.0 & 195.0 \\ 
11 & 295.2 & 295.2 & 0 & 105.0 & 190.0 \\ 
12 & 344.4 & 344.4 & 0 & 110.0 & 185.0 \\ 
\enddata


\tablenotetext{a}{The first row gives the semi-major, semi-minor axis,
and orientation of the inner ellipse of annulus 1.  All other values
refer to the outer boundary of the corresponding annulus.}
\tablenotetext{b}{Semi-major axis of ellipse.}
\tablenotetext{c}{Semi-minor axis of ellipse.}
\tablenotetext{d}{P.A. of major axis of ellipse measured N through E.}
\tablenotetext{e}{P.A. of beginning angle(s) of partial ellipse(s).}
\tablenotetext{f}{P.A. of ending angle(s) of partial ellipse(s).}

\end{deluxetable} 

\clearpage

\begin{deluxetable}{ccccccccc}
\footnotesize
\rotate
\tablecaption{Spectral fits to the integrated intracluster emission imaged 
on chip S3 in the 0.75--8~keV band. \label{tbl-2}}
\tablewidth{0pt}
\tablehead{
\colhead{Model} &
\colhead{$N_{\rm H}$\tablenotemark{a}} &
\colhead{$kT_{\rm l}$\tablenotemark{b}} &
\colhead{$Z_{\rm l}$\tablenotemark{c}} &
\colhead{Norm$_{\rm l}$\tablenotemark{d}} &
\colhead{$kT_{\rm h}$\tablenotemark{e}} &
\colhead{$Z_{\rm h}$\tablenotemark{f}} &
\colhead{Norm$_{\rm h}$\tablenotemark{g}} &
\colhead{$\chi^{2}$ (d.o.f.)} \\
\colhead{} &
\colhead{($10^{21}$ cm$^{-2}$)} &
\colhead{(keV)} &
\colhead{($Z_{\odot}$)} &
\colhead{($\frac{10^{-14}}{4 \pi d_{\rm A}^{2}} \int n_{\rm e} n_{\rm H} {\rm d}V$)} &
\colhead{(keV)} &
\colhead{($Z_{\odot}$)} &
\colhead{($\frac{10^{-14}}{4 \pi (d_{\rm A}(1+z))^{2}} \int n_{\rm e} n_{\rm H} {\rm d}V$)} &
\colhead{}}
\startdata
1 & $2.815^{+0.075}_{-0.075}$ & ... & ... & ... & $8.26^{+0.37}_{-0.37}$ & 
	$0.346^{+0.051}_{-0.050}$ & $(2.344^{+0.031}_{-0.031}) \times 
	10^{-2}$ & 736.2 (489) \\
2 & $3.39^{+0.29}_{-0.28}$ & $0.67^{+0.14}_{-0.04}$ & $1.0^{\rm fixed}$ &
	$(3.1^{+1.6}_{-1.2}) \times 10^{-4}$ & $7.66^{+0.44}_{-0.44}$
	& $0.336^{+0.048}_{-0.047}$ & $(2.407^{+0.052}_{-0.040})
	\times 10^{-2}$ & 709.4 (487) \\
\enddata


\tablenotetext{a}{Measured column density.}
\tablenotetext{b}{Low temperature thermal component intended to
represent spatially uniform Galactic emission.}
\tablenotetext{c}{Metal abundance of low temperature thermal component.}
\tablenotetext{d}{Normalization of the low temperature thermal
component.  $d_{\rm A}$ refers to the effective distance of this
component.  The units are c.g.s.}
\tablenotetext{e}{High temperature thermal component intended to
represent cluster emission.}
\tablenotetext{f}{Metal abundance of high temperature thermal component.}
\tablenotetext{g}{Normalization of the high temperature thermal
component.  The units are c.g.s.}

\tablecomments{The errors given here are not strictly correct because
both models provide unacceptable descriptions of the data (e.g.,
Lampton, Margon, \& Bowyer 1976).}

\end{deluxetable} 

\clearpage

\begin{deluxetable}{ccccccccc}
\footnotesize
\rotate
\tablecaption{Spectral fits to the observed emission from the annuli
with a two-component mekal model and constant X-ray absorption in the
0.75--8~keV band. \label{tbl-3}}
\tablewidth{0pt}
\tablehead{
\colhead{Annulus\tablenotemark{a}} &
\colhead{$N_{\rm H}$\tablenotemark{b}} & 
\colhead{$kT_{\rm l}$\tablenotemark{c}} &
\colhead{$Z_{\rm l}$\tablenotemark{d}} &
\colhead{Norm$_{\rm l}$\tablenotemark{e}} &
\colhead{$kT_{\rm h}$\tablenotemark{f}} &
\colhead{$Z_{\rm h}$\tablenotemark{g}} &
\colhead{Norm$_{\rm h}$\tablenotemark{h}} &
\colhead{$\chi^{2}$ (d.o.f.)\tablenotemark{i}} \\
\colhead{} &
\colhead{($10^{21}$ cm$^{-2}$)} &
\colhead{(keV)} &
\colhead{($Z_{\odot}$)} &
\colhead{($\frac{10^{-14}}{4 \pi d_{\rm A}^{2}} \int n_{\rm e} n_{\rm H} {\rm d}V$)} &
\colhead{(keV)} &
\colhead{($Z_{\odot}$)} &
\colhead{($\frac{10^{-14}}{4 \pi (d_{\rm A}(1+z))^{2}} \int n_{\rm e} n_{\rm H} {\rm d}V$)} &
\colhead{}}
\startdata
1 & $3.109^{+0.083}_{-0.086}$ & $0.593^{+0.07}_{-0.11}$ & $1.0^{\rm fixed}$ & $(5.6^{+1.7}_{-1.4}) \times 10^{-7}$ & $5.27^{+0.40}_{-0.40}$ & $0.80^{+0.20}_{-0.18}$ & $(1.347^{+0.065}_{-0.064}) \times 10^{-3}$ & 2083.7 (2021) \\
2 & ... & ... & ... & $(5.8^{+1.7}_{-1.5}) \times 10^{-7}$ & $5.77^{+0.78}_{-0.53}$ & $0.68^{+0.26}_{-0.23}$ & $(8.00^{+0.47}_{-0.48}) \times 10^{-4}$ & ... \\
3 & ... & ... & ... & $(6.8^{+2.0}_{-1.7}) \times 10^{-7}$ & $6.45^{+0.85}_{-0.74}$ & $0.72^{+0.36}_{-0.30}$ & $(6.93^{+0.48}_{-0.49}) \times 10^{-4}$ & ... \\
4 & ... & ... & ... & $(1.01^{+0.30}_{-0.26}) \times 10^{-6}$ & $7.7^{+1.3}_{-1.1}$ & $0.55^{+0.29}_{-0.27}$ & $(7.94^{+0.45}_{-0.46}) \times 10^{-4}$ & ... \\
5 & ... & ... & ... & $(1.44^{+0.43}_{-0.37}) \times 10^{-6}$ & $6.69^{+0.99}_{-0.88}$ & $0.42^{+0.25}_{-0.23}$ & $(8.19^{+0.47}_{-0.47}) \times 10^{-4}$ & ... \\
6 & ... & ... & ... & $(3.9^{+1.2}_{-1.0}) \times 10^{-6}$ & $7.6^{+1.0}_{-0.7}$ & $0.41^{+0.20}_{-0.18}$ & $(1.606^{+0.065}_{-0.065}) \times 10^{-3}$ & ... \\
7 & ... & ... & ... & $(9.4^{+2.8}_{-2.4}) \times 10^{-6}$ & $7.11^{+0.62}_{-0.57}$ & $0.32^{+0.13}_{-0.13}$ & $(2.740^{+0.083}_{-0.086}) \times 10^{-3}$ & ... \\
8 & ... & ... & ... & $(1.86^{+0.55}_{-0.47}) \times 10^{-5}$ & $7.63^{+0.80}_{-0.61}$ & $0.34^{+0.13}_{-0.13}$ & $(3.242^{+0.096}_{-0.095}) \times 10^{-3}$ & ... \\
9 & ... & ... & ... & $(2.31^{+0.69}_{-0.59}) \times 10^{-5}$ & $8.5^{+1.1}_{-1.0}$ & $0.30^{+0.20}_{-0.18}$ & $(2.097^{+0.082}_{-0.082}) \times 10^{-3}$ & ... \\
10 & ... & ... & ... & $(1.78^{+0.53}_{-0.45}) \times 10^{-5}$ & $7.8^{+1.7}_{-1.3}$ & $0.02^{+0.30}_{-0.02}$ & $(1.062^{+0.037}_{-0.064}) \times 10^{-3}$ & ... \\
11 & ... & ... & ... & $(1.76^{+0.52}_{-0.45}) \times 10^{-5}$ & $7.6^{+1.9}_{-1.3}$ & $0.65^{+0.57}_{-0.46}$ & $(5.98^{+0.58}_{-0.57}) \times 10^{-4}$ & ... \\
12 & ... & ... & ... & $(1.84^{+0.55}_{-0.47}) \times 10^{-5}$ & $8.7^{+5.6}_{-2.5}$ & $0.22^{+0.75}_{-0.22}$ & $(4.66^{+0.42}_{-0.57}) \times 10^{-4}$ & ... \\
\enddata


\end{deluxetable} 

\clearpage

\setlength{\parindent}{0pt}

$^{a}$Defined in Table~\ref{tbl-1} and Figure~\ref{fig6}. \\
$^{b}$Measured column density. \\
$^{c}$Low temperature thermal component intended to represent 
spatially uniform Galactic emission. \\
$^{d}$Metal abundance of low temperature thermal component. \\
$^{e}$Normalization of the low temperature thermal component.  $d_{\rm
A}$ refers to the effective distance of this component.  The units are
c.g.s. \\
$^{f}$High temperature thermal component intended to represent
cluster emission. \\
$^{g}$Metal abundance of high temperature thermal component. \\
$^{h}$Normalization of the high temperature thermal component.  The
units are c.g.s. \\
$^{i}$This is the value of $\chi^{2}$ (d.o.f.) for the fit to all of
the annuli. \\

\clearpage

\begin{deluxetable}{ccccccccc}
\footnotesize
\rotate
\tablecaption{Deprojected temperatures, abundances, and emissivities
from spectral fits in the 0.75--8~keV band. \label{tbl-4}}
\tablewidth{0pt}
\tablehead{
\colhead{Shell\tablenotemark{a}} &
\colhead{$N_{\rm H}$\tablenotemark{b}} & 
\colhead{$kT_{\rm l}$\tablenotemark{c}} &
\colhead{$Z_{\rm l}$\tablenotemark{d}} &
\colhead{Norm$_{\rm l}$\tablenotemark{e}} &
\colhead{$kT_{\rm h}$\tablenotemark{f}} &
\colhead{$Z_{\rm h}$\tablenotemark{g}} &
\colhead{Norm$_{\rm h}$\tablenotemark{h}} &
\colhead{$\chi^{2}$ (d.o.f.)} \\
\colhead{} &
\colhead{($10^{21}$ cm$^{-2}$)} &
\colhead{(keV)} &
\colhead{($Z_{\odot}$)} &
\colhead{($\frac{10^{-14}}{4 \pi d_{\rm A}^{2}} \int n_{\rm e} n_{\rm H} {\rm d}V$)} &
\colhead{(keV)} &
\colhead{($Z_{\odot}$)} &
\colhead{($\frac{10^{-14}}{4 \pi (d_{\rm A}(1+z))^{2}} \int n_{\rm e} n_{\rm H} {\rm d}V$)} &
\colhead{}} 
\startdata
1 & $3.109$ & $0.593$ & $1.0$ & $5.6 \times 10^{-7}$ & $4.91^{+0.62}_{-0.56}$ & $0.89^{+0.35}_{-0.30}$ & $(9.97^{+0.83}_{-0.82}) \times 10^{-3}$ & 2084.2 (2057) \\
2 & ... & ... & ... & $5.8 \times 10^{-7}$ & $4.9^{+1.7}_{-1.1}$ & $0.70^{+0.76}_{-0.55}$ & $(3.57^{+0.63}_{-0.61}) \times 10^{-3}$ & 1888.6 (1888) \\
3 & ... & ... & ... & $6.8 \times 10^{-7}$ & $5.6^{+1.4}_{-1.1}$ & $0.95^{+0.86}_{-0.60}$ & $(3.98^{+0.65}_{-0.65}) \times 10^{-3}$ & 1760.5 (1759) \\
4 & ... & ... & ... & $1.01 \times 10^{-6}$ & $9.2^{+10.4}_{-2.5}$ & $0.66^{+0.81}_{-0.66}$ & $(3.17^{+0.55}_{-0.41}) \times 10^{-3}$ & 1649.1 (1637) \\
5 & ... & ... & ... & $1.44 \times 10^{-6}$ & $5.4^{+2.2}_{-1.3}$ & $0.45^{+0.68}_{-0.45}$ & $(2.01^{+0.33}_{-0.32}) \times 10^{-3}$ & 1531.6 (1507) \\
6 & ... & ... & ... & $3.9 \times 10^{-6}$ & $8.5^{+3.5}_{-2.0}$ & $0.59^{+0.64}_{-0.59}$ & $(2.27^{+0.25}_{-0.25}) \times 10^{-3}$ & 1413.7 (1381) \\
7 & ... & ... & ... & $9.4 \times 10^{-6}$ & $6.7^{+1.3}_{-1.0}$ & $0.29^{+0.29}_{-0.26}$ & $(3.85^{+0.25}_{-0.26}) \times 10^{-3}$ & 1248.5 (1195) \\
8 & ... & ... & ... & $1.86 \times 10^{-5}$ & $7.0^{+1.1}_{-0.9}$ & $0.39^{+0.23}_{-0.22}$ & $(5.12^{+0.26}_{-0.26}) \times 10^{-3}$ & 937.0 (949) \\
9 & ... & ... & ... & $2.31 \times 10^{-5}$ & $9.3^{+3.4}_{-2.0}$ & $0.32^{+0.47}_{-0.32}$ & $(4.15^{+0.34}_{-0.33}) \times 10^{-3}$ & 626.2 (673) \\
10 & ... & ... & ... & $1.78 \times 10^{-5}$ & $7.6^{+3.6}_{-2.0}$ & $< 0.23$ & $(4.29^{+0.27}_{-0.27}) \times 10^{-3}$ & 395.3 (435) \\
11 & ... & ... & ... & $1.76 \times 10^{-5}$ & $6.9^{+3.8}_{-2.2}$ & $1.3^{+2.3}_{-1.2}$ & $(2.26^{+0.61}_{-0.62}) \times 10^{-3}$ & 225.3 (260) \\
12 & ... & ... & ... & $1.84 \times 10^{-5}$ & $8.7^{+5.0}_{-2.4}$ & $0.22^{+0.75}_{-0.22}$ & $(6.07^{+0.52}_{-0.74}) \times 10^{-3}$ & 100.6 (112) \\
\enddata


\end{deluxetable} 

\clearpage

\setlength{\parindent}{0pt}

$^{a}$Each shell is the volume between two prolate spheroids or
spheres which project to give the annuli defined in Table~\ref{tbl-1}
and Figure~\ref{fig6}. \\
$^{b}$Measured column density. \\
$^{c}$Low temperature thermal component intended to represent 
spatially uniform Galactic emission. \\
$^{d}$Metal abundance of low temperature thermal component. \\
$^{e}$Normalization of the low temperature thermal component.  $d_{\rm
A}$ refers to the effective distance of this component.  The units are
c.g.s. \\
$^{f}$High temperature thermal component intended to represent
cluster emission. \\
$^{g}$Metal abundance of high temperature thermal component. \\
$^{h}$Normalization of the high temperature thermal component.  Note
that this parameter refers to volume contained between the two prolate
spheroids.  The units are c.g.s. \\

\clearpage 

\begin{deluxetable}{ccccccc}
\footnotesize
\rotate
\tablecaption{Properties of the intracluster gas for each of the
spheroidal or spherical shells. \label{tbl-5}}
\tablewidth{0pt}
\tablehead{
\colhead{Mean Radius\tablenotemark{a}} &
\colhead{Volume} & 
\colhead{$F_{\rm X}$\tablenotemark{b}} &
\colhead{$L_{\rm X}$\tablenotemark{c}} &
\colhead{$n_{\rm e}$\tablenotemark{d}} &
\colhead{$P_{\rm g}$\tablenotemark{e}} &
\colhead{$\tau_{\rm c}$\tablenotemark{f}} \\
\colhead{(kpc)} &
\colhead{($10^{72}$ cm$^{3}$)} &
\colhead{($10^{-11}$ erg cm$^{-2}$ s$^{-1}$)} &
\colhead{($10^{44}$ erg s$^{-1}$)} &
\colhead{($10^{-2}$ cm$^{-3}$)} &
\colhead{($10^{-10}$ erg cm$^{-3}$)} &
\colhead{($10^{10}$ yr)}}
\startdata
72.0 & $0.0144$ & $1.2$ & $1.4$ & $3.0$ & $2.3$ & $0.078$ \\
79.9 & $0.0164$ & $0.42$ & $0.46$ & $1.7$ & $1.3$ & $0.15$ \\
87.3 & $0.0204$ & $0.53$ & $0.60$ & $1.6$ & $1.4$ & $0.15$ \\
95.7 & $0.0328$ & $0.45$ & $0.53$ & $1.1$ & $1.6$ & $0.32$ \\
105.9 & $0.0503$ & $0.23$ & $0.25$ & $0.72$ & $0.62$ & $0.39$ \\
120.7 & $0.0900$ & $0.31$ & $0.36$ & $0.57$ & $0.77$ & $0.61$ \\
148.6 & $0.274$ & $0.45$ & $0.52$ & $0.43$ & $0.46$ & $0.77$ \\
195.0 & $0.708$ & $0.63$ & $0.72$ & $0.30$ & $0.34$ & $1.1$ \\
260.0 & $1.68$ & $0.55$ & $0.64$ & $0.18$ & $0.27$ & $2.2$ \\
334.3 & $2.77$ & $0.49$ & $0.56$ & $0.14$ & $0.17$ & $2.7$ \\
408.6 & $4.13$ & $0.35$ & $0.41$ & $0.084$ & $0.093$ & $3.0$ \\
482.9 & $-$ & $0.85$ & $0.89$ & $-$ & $-$ & $-$ \\
\enddata


\end{deluxetable} 

\clearpage

\setlength{\parindent}{0pt}

$^{a}$For each ellipsoidal shell, the mean radius is the average of the 
semi-major and semi-minor axis. \\
$^{b}$Observed 0.5--10~keV flux. \\
$^{c}$Unobscured 2--10~keV rest-frame luminosity, calculated assuming
a luminosity distance of $d_{\rm L} = 346.4$~Mpc. \\
$^{d}$Derived electron density. \\
$^{e}$Derived gas pressure. \\
$^{f}$Derived cooling time. \\

\clearpage 

\begin{deluxetable}{cccccc}
\footnotesize
\tablecaption{Spectral fits to the extended filaments, bands of X-ray
emission near the nucleus, and limb-brightened regions of the cavity,
in the 0.5--10~keV band. \label{tbl-6}}
\tablewidth{0pt}
\tablehead{
\colhead{Region} &
\colhead{$N_{\rm H}$\tablenotemark{a}} & 
\colhead{$kT$\tablenotemark{b}} &
\colhead{$Z$\tablenotemark{c}} &
\colhead{Norm\tablenotemark{d}} &
\colhead{$\chi^{2}$ (d.o.f.)} \\
\colhead{} &
\colhead{($10^{21}$ cm$^{-2}$)} &
\colhead{(keV)} &
\colhead{($Z_{\odot}$)} &
\colhead{($\frac{10^{-14}}{4 \pi (d_{\rm A}(1+z))^{2}} \int n_{\rm e} n_{\rm H} {\rm d}V$)} &
\colhead{}}
\startdata
A & $3.57^{+0.21}_{-0.20}$ & $5.59^{+0.72}_{-0.54}$ & $0.70^{+0.22}_{-0.19}$ & $(9.92^{+0.64}_{-0.61}) \times 10^{-4}$ & 152.7 (152) \\
B & $2.78^{+0.22}_{-0.18}$ & $6.19^{+0.84}_{-0.90}$ & $0.63^{+0.22}_{-0.20}$ & $(8.70^{+0.60}_{-0.51}) \times 10^{-4}$ & 204.8 (148) \\
C & $3.15^{+0.24}_{-0.22}$ & $3.89^{+0.42}_{-0.39}$ & $0.84^{+0.27}_{-0.23}$ & $(6.35^{+0.63}_{-0.56}) \times 10^{-4}$ & 117.9 (111) \\
D & $3.63^{+0.32}_{-0.29}$ & $4.45^{+0.71}_{-0.56}$ & $0.88^{+0.39}_{-0.31}$ & $(4.08^{+0.48}_{-0.43}) \times 10^{-4}$ & 64.1 (71) \\
E & $3.11^{+0.29}_{-0.25}$ & $3.79^{+0.45}_{-0.41}$ & $0.80^{+0.28}_{-0.25}$ & $(5.18^{+0.57}_{-0.49}) \times 10^{-4}$ & 89.3 (91) \\
F & $3.24^{+0.33}_{-0.28}$ & $3.97^{+0.56}_{-0.54}$ & $0.76^{+0.37}_{-0.29}$ & $(3.79^{+0.49}_{-0.43}) \times 10^{-4}$ & 49.7 (70) \\
G & $3.53^{+0.35}_{-0.31}$ & $4.48^{+0.69}_{-0.53}$ & $0.97^{+0.45}_{-0.36}$ & $(3.69^{+0.46}_{-0.43}) \times 10^{-4}$ & 87.5 (72) \\
H & $2.82^{+0.26}_{-0.24}$ & $7.0^{+1.7}_{-1.1}$ & $0.97^{+0.52}_{-0.42}$ & $(4.39^{+0.44}_{-0.43}) \times 10^{-4}$ & 113.7 (87) \\
I & $3.76^{+0.31}_{-0.28}$ & $4.43^{+0.55}_{-0.46}$ & $1.39^{+0.52}_{-0.41}$ & $(4.31^{+0.53}_{-0.50}) \times 10^{-4}$ & 85.9 (82) \\
J & $3.36^{+0.22}_{-0.21}$ & $6.58^{+0.96}_{-0.94}$ & $0.64^{+0.28}_{-0.24}$ & $(7.66^{+0.53}_{-0.51}) \times 10^{-4}$ & 141.6 (132) \\
\enddata


\tablenotetext{a}{Measured column density.}
\tablenotetext{b}{Temperature.}
\tablenotetext{c}{Metal abundance.}
\tablenotetext{d}{Normalization of the mekal model.  The units are c.g.s.}

\end{deluxetable} 

\clearpage 

\end{document}